\documentclass[a4paper]{article}
\usepackage{amssymb}
\usepackage{amsmath}
\newcommand{\be}{\begin{equation}}
\newcommand{\ee}{\end{equation}}

\begin{document}

\title{Nonlinear deterministic-chaotic collapse model - preliminaries, philosophy, locality}

\author{Tam\'as Geszti\\Department of Physics of Complex Systems \\ E\"otv\"os University, 
Budapest, Hungary}

\maketitle

\footnote{email: geszti@elte.hu}

\begin{abstract}
This is an extended discussion of Ref. \cite{jphysa18}, presenting a nonlinear dynamical model of quantum collapse, with randomness emerging from self-generated noise. Here we focus on a few issues: 1) the way chaos theory explains ``deterministic but unpredictable'' as a generic feature of nonlinear dynamics; 2) a new argument about why Bell-CHSH and GHZ experiments confirm quantum mechanics; 3) a discussion of why the heating effect predicted by CSL-type theories is not expected to happen according to our approach; 4) in looking for direct-product solutions of our nonlinear Von Neumann equation, we use a parametrization different from that used in Ref. \cite{jphysa18}, allowing better insight to the locality issue. 
\end{abstract}

\section{Introduction: unpredictable determinism}

Quantum measurement is a nonlinear phenomenon: an entering wave is getting separated into a superposition of partial waves, each facing a detector; finally, one of the detectors, apparently randomly chosen, gives a signal, the others not. That requires {\sl pumping} weights of a superposition from one partial wave into another; a nonlinear process, since in linear evolution those weights remain constant. On the contrary, quantum phenomena are routinely described by the linear Schr\"odinger equation, offering an extremely accurate description of whatever happens before the measurement.

That contradiction is obviously violating our longstanding belief that physics is based on measurements, observing objective properties of the world around us, furnishing data independent of the act of measurement. A successful escape is the Copenhagen interpretation: a special status is assigned to the measurement process, meant to offer random data at probabilities determined by the solutions of the 
Schr\"odinger equation through Born's rule.

As recognized by Einstein \cite{ei,EPR}, whether measurement results reflect properties of the observed system independent of the measuring apparatus (``reality''), is open to experimental tests by observing detector-detector correlations. Further analysis focused experimental tests on two-particle correlations \cite{bell,chsh} with a flexible control on detector setting; later extended to three-particle correlations \cite{ghz}. An enormous amount of experimental evidence accumulated since that time \cite{hensenhanson} demonstrates that the expectations are {\sl not} satisfied: measurement outcomes are not determined by any known or unknown properties of the incoming particles (``local reality''); rather, individual outcomes emerge in the process of measurement; only their statistics can be predicted through Born's rule. This is in clear contradiction with our classicality-based prejudices about apparatus-independent reality, and suggests we should accept that the actual state of the apparatus, including all its microscopic details, is part of the reality displayed in measurement outcomes.

On the contrary, this is not in contradiction with full determinism, as we learned in the 1970's from the rise of nonlinear chaotic dynamics \cite{chaos}; the trivial example of which is throwing dice. The key idea is {\sl sensitivity to initial conditions,} making the evolution unpredictable, even if deterministic, since one is unable to control initial conditions accurately enough to decide between markedly different outcomes. As foreseen by Lyapunov's groundbreaking treatment of nonlinear differential equations, and subsequently identified on low-dimension models of chaos \cite{chaos}, this is a decisive property of macroscopic bodies, like a particle detector, deliberately prepared in a metastable initial state \cite{vankam,NSgeszti}.

All that points into the direction of looking for a nonlinear dynamical equation, reproducing the collapse phenomenology, including Born's rule. The treatment should focus on finding a description for the events of pumping; the route to Born's rule is then open through Pearle's ``gambler's ruin'' mechanism \cite{gambler}. 

Generically, nonlinear dynamics carries the danger of superluminal signaling \cite{gisinhelv,signaling}; however, a shortcut is offered by the observation that fully reproducing quantum predictions implies no signaling \cite{brunner}.

Taking that route, in Ref. \cite{jphysa18} we presented a deterministic evolution equation for the density matrix, containing a minimal nonlinear extension of the Von Neumann equation. As briefly summarized below, straightforward analysis of that equation is capable to fully reproduce the collapse phenomenology of quantum measurement, with random outcomes emerging from nonlinear dynamics, at probabilities determined by Born's rule. That would resolve the signaling issue, with one contradiction remaining: our treatment excludes signaling through statistics, not in individual events, as explicitely seen in the equations. Although quantum predictions refer to statistics only, individual events are part of the reality around us - this is a logical catch that our treatment is unable to avoid.
 
Our treatment is close to the widely studied model class called CSL \cite{csl}, with important differences: 
those models postulate an omnipresent external noise, fundamentally connected to the actual quantum state through Born's rule, whereas we consider self-generated noise emerging in  detector dynamics, reproducing Born's rule only in measurement situations. Analysing non-measurement scenarios on the quantum-classical border, amply discussed in terms of CSL-type collapse models, may require a different approach; as mentioned in Ref. \cite{jphysa18}, a possible explanation of $1/f$ noise is one of the obvious scopes of further research.

\section{The nonlinear equation}

As described in detail in Ref. \cite{jphysa18}, we are looking for strictly deterministic evolution of an individual system, composed of a microsystem and a delocalized set of detectors. Evolution should be nonlinear, which means that the Hamiltonian $\hat H$ should depend on the actual  quantum state of the composed system. Our first task is to specify the form of that dependence.

The requirement of gauge invariance excludes linear dependence on the state vector $\bf|\Psi\rangle$; the obvious alternative is a Hamiltonian $\hat H(\hat{\boldsymbol\varrho})$ depending on the density matrix $\hat{\boldsymbol\varrho}$. Since nothing physically disappears during collapse, only weights of a superposition - constant in linear dynamics - are pumped by nonlinearity from one quantum state into another; their sum remaining unchanged. Accordingly, the evolution should be unitary, and $\hat H$ Hermitian. To account for the accuracy of linear dynamics on the microscopic level, we are looking for a minimal nonlinearity, with $\hat H(\hat{\boldsymbol\varrho})$ depending linearly on $\hat{\boldsymbol\varrho}$. The obvious choice is then to add nonlinearity in the form of a single Kraus operator $\hat M \hat{\boldsymbol\varrho} \hat M^\dagger$ to the linear Hamiltonian $\hat H_0$.  Accordingly, we are going to study evolution  governed by a nonlinear Von Neumann equation
\be\begin{split}\label{vNnonlin}
 \partial_t \hat{\boldsymbol\varrho} = &
      -\frac{i}{\hbar}\left[\hat H(\hat{\boldsymbol\varrho}),\hat{\boldsymbol\varrho}\right]\\
      &= -\frac{i}{\hbar}\left[\left(\hat H_0+\zeta \hat M \hat{\boldsymbol\varrho}\hat M^\dagger\right),\hat{\boldsymbol\varrho}\right].
\end{split}\ee

Following Ref. \cite{jphysa18}, we focus on typical measurement situations, and describe collapse as pumping weights among terms of a superposition. Quantum measurement is done by a set of remote detectors. It starts by a Stern-Gerlach-type separation process of incoming particles, open to control by changing beamsplitter orientations and detector positions. Separation results in a superposition of the form $\sum_k c_k|k\rangle$, expanded on an orthogonal basis of non-overlapping partial waves. Each basis state is defined by a different combination of those detectors which are hit by particles, and those left quiet - a distinctive feature of measurement situations, shared by Stern-Gerlach, Bell-CHSH and GHZ\footnote{As an example, in a Bell measurement, in basis state $|k\rangle=|+-\rangle$ of two entangled particles from a common source, left-going particle is directed towards detector left-up, right-going particle towards detector right-down. It should be noted that these basis states are not what is called ``Bell states''.}. Our derivation below demonstrates that this is sufficient to arrive at Born's rule; we conjecture it is necessary as well, and in non-measurement situations non-Born statistical features may emerge.

Interaction with the detectors at time $t=0$, through fast linear quantum dynamics, creates an entangled state of particles and detectors in the form
\be\label{Psi}
{\bf|\Psi\rangle}~=~\sum_k c_k|k\rangle|\Phi_k\rangle,
\ee
which becomes the initial state for subsequent temporal changes concluding in collapse.  $|\Phi_k(t)\rangle$ denotes a multi-detector pure state vector, with local environments of each detector included. 

The density matrix corresponding to state vector (\ref{Psi}) takes the form 
\be\label{dnstmtrx}
\hat{\boldsymbol\varrho}~=~{\bf|\Psi\rangle\langle\Psi|}~=~
       \sum_{k,l} ~ |k\rangle\langle l|~c_kc_l^*~ \hat R_{kl}(t),
\ee
with
\be\label{Rdef}
\hat R_{kl}~=~|\Phi_k\rangle\langle\Phi_l|.
\ee
In coordinate representation, the operator $\hat R_{kl}$ appears as a Hermitian matrix $R_{kl}( x, x';t)=R_{lk}( x', x;t)^*$, acting on the configuration space of all detectors, including their local environments.

The weight of state $|k\rangle|\Phi_k\rangle$ in the superposition is given by
\be\label{weight}
w_k(t) = |c_k|^2~Tr_x\hat R_{kk}(t) = |c_k|^2~\langle\Phi_k(t)|\Phi_k(t)\rangle.
\ee
Unlike in Ref. \cite{jphysa18}, now we postulate $|c_k|^2 = const$, and follow the evolution of $w_k(t)$ - pumping from one state to the other, ending in collapse - through the nonlinear dynamics of $\hat R_{kl}(t)$. Apart from the initial normalization
\be\label{initnorm}
w_k(t=0)=|c_k|^2~~\Longrightarrow~~Tr_x\hat R_{kk}(t=0)=1 ~~~\forall k, 
\ee
the possible initial states of the macroscopic measurement apparatus are of enormous variety, serving as the multi-local hidden parameters mentioned in the Introduction.

Since we expect nonlinearity to dominate dynamics of the macroscopic apparatus, we postulate the Kraus operator $\hat M$ to act on the configuration space coordinates, not on indices $k$. Then from Eqs. (\ref{weight}) and (\ref{vNnonlin}) we immediately obtain
\be\label{pumpKraus}
\dot w_k~=~-i \frac{\zeta}{\hbar}~\sum_m |c_k|^2|c_m|^2 
~Tr_x \left(\hat M \hat R_{km} \hat M^\dagger \hat R_{mk} -
\hat R_{km} \hat M \hat R_{mk} \hat M^\dagger \right).       
\ee
By cyclic invariance of the trace, this cancels to 0 if $\hat M$ is self-adjoint, so to describe pumping, we have to postulate a non-Hermitian $\hat M=\hat M_1+i\hat M_2$, with $\hat M_1$ and $\hat M_2$ non-zero Hermitian operators. Then dropping the non-pumping combinations $\hat M_1 \hat{\boldsymbol\varrho}\hat M_1$ and $\hat M_2 \hat{\boldsymbol\varrho}\hat M_2$, what remains is a term $i\zeta(\hat M_1 \hat{\boldsymbol\varrho}\hat M_2 - \hat M_2 \hat{\boldsymbol\varrho}\hat M_1)$ in the Hamiltonian. 
 
   Trying to give some physical identity to $\hat M_1$ and $\hat M_2$, while keeping the full nonlinear Hamiltonian scalar, by intuition we choose $\hat x$ and $\hat p$ as configuration-space vectors, and the whole combination a scalar product over the configuration space. Accordingly, in the following we use Equation (\ref{vNnonlin}) in the final form
\be\label{vNnonlinfinal}
 \partial_t \hat{\boldsymbol\varrho}= -\frac{i}{\hbar} \left[ \hat H_0,\hat{\boldsymbol\varrho}\right]  + \frac{\zeta}{\hbar} 
  \left[\left( \hat x \hat{\boldsymbol\varrho}\hat p -  \hat p \hat{\boldsymbol\varrho} \hat x \right) ,\hat{\boldsymbol\varrho}\right]
\ee
where $\hat H_0$ describes linear evolution of the incoming particles and all detectors interacting with them, including  environments of the whole measurement setup.\footnote{Rejecting the non-pumping terms is not fully trivial, since the model should be tested in non-measurement situations too. In that context, we notice that rejecting $ \hat x \hat{\boldsymbol\varrho} \hat x $ is trivial, since that term would violate translation invariance. On the contrary,  a term $ \hat p \hat{\boldsymbol\varrho} \hat p $ would add a kind of kinetic energy with a rest mass depending on the quantum state to the Hamiltonian, which is not absurd; anyway, that would introduce a length scale into the framework.} 

The constant $\zeta$ characterizes the strength of nonlinearity. Since the combination $\hat x \hat{\boldsymbol\varrho}\hat p - \hat p \hat{\boldsymbol\varrho} \hat x$ has the dimension of action, measured in quantum units by the factor $\hbar^{-1}$, $\zeta$ is of dimensionality $t^{-1}$. The immediate insight gained therefrom is to connect the quantum-classical border to time scales: for microscopic systems, moving ${\cal O}(1)$ amount of action, quantum phenomena go to the end without noticeable signature of nonlinearity; macroscopic amount of action makes nonlinearity immediately dominant. In Ref. \cite{jphysa18},  $\zeta \approx  10^{-10} ~s^{-1}$, $\zeta^{-1} \approx 1/2 ~yr$ is proposed as a rough estimate, based on known parameters of an avalanche photon detector.

\section{The route to Born's rule}

In what follows, we use the interaction picture and look for a solution of Eq. (\ref{vNnonlinfinal}) in the form of Eq. (\ref{dnstmtrx}). As a preparation, we evaluate the mean value of a configuration-space operator $\hat A_x$ acting on the  configuration space coordinates, not on indices $k$; with the result
\be\label{mean}
\langle \hat A_x \rangle  ~=~\sum_k |c_k|^2~ \langle\Phi_k|\hat A_x |\Phi_k\rangle
~=~\sum_k w_k~ \langle \hat A_x \rangle_k ,
 \ee
where
\be\label{meanink}
 \langle \hat A_x \rangle_k~=~\frac{\langle\Phi_k| A_x |\Phi_k\rangle}{\langle\Phi_k|\Phi_k\rangle}
   ~=~\frac{Tr_x\left(\hat R_{kk}\hat A_x\right)}{Tr_x~\hat R_{kk}}
\ee
is the mean value of $\hat A_x$ in basis state $k$.

Considering that the vector operators $\hat x$ and $\hat p$ are configuration-space operators as defined above, and using Eq. (\ref{meanink}), in straightforward steps one arrives at the system of equations
\be\label{nondiag} 
 \frac{\partial \hat R_{kl}}{\partial t}=~\frac{\zeta}{\hbar}~\sum_m w_m 
\left(\langle\hat p\rangle_m \cdot \{\hat x,\hat R_{kl}\} 
- \langle\hat x\rangle_m\cdot\{\hat p,\hat R_{kl}\}\right)
\ee
where $\{...,...\}$ denotes the anticommutator. 

To calculate the pumping rates, we focus on $l=k$. Starting from Eq. (\ref{weight}),
using  cyclic invariance of the trace, one finally obtains the pumping rates in the form
\be\label{pump}
\dot w_k~=~\zeta~ \sum_m w_k w_m A_{km},
\ee
where 
\be\label{toandfro}
A_{km}(t)~=~\frac{2}{\hbar}~  \Big(\langle\hat x\rangle_k \cdot \langle\hat p\rangle_m
             ~-~ \langle\hat x\rangle_m \cdot \langle\hat p\rangle_k\Big).
\ee
Eqs. (\ref{pump}) with (\ref{toandfro}) constitute a system of balance equations, with  the obvious antisymmetry property
\be\label{antisymm}
A_{km}=-A_{mk},
\ee 
granting $\sum_k\dot w_k=0$, i.e., conservation of the total weight, traced back to self-adjointness of the Hamiltonian.  

It is the dimensionless quantities $A_{km}$ defined by Eq. (\ref{toandfro}) which - although obeying deterministic dynamics under control of Hamiltonian $\hat H(\hat{\boldsymbol\varrho})$ - behave like a random noise, since the dynamics of metastable detectors is chaotic, and thereby sensitive to details of the initial states of the detectors, impossible to control. That quasi-randomness of $A_{km}(t)$ makes the process described by Eq. (\ref{pump}) similar to a stochastic game of the ``gambler's ruin'' type \cite{gambler}.

Eq. (\ref{pump}) is immediately seen to describe some kind of dynamics in which weights $w_k$ are vanishing one after the other; once vanished, that weight never reappears; finally, one ``winner'' remains alive at $w_m=1$, all the rest disappearing, in accordance with the ``gambler's ruin'' model. Randomness in quantum measurements emerges from chaotic nonlinear dynamics of the macroscopic detectors biased to a metastable state, giving $A_{km}(t)$ the character of a noise, self-generated by deterministic dynamics. 

As discussed in detail in Ref. \cite{jphysa18}, the no-drift property $\langle A_{km}(t) \rangle=0$, equivalent to the ``fair play'' or ``martingale'' property in the language of games, is sufficient to assure that the probability of a given outcome $m$ is equal to the initial weight $|c_m|^2$, which is Born's rule. By modeling $A_{km}(t)$ as white noise, all that can be followed in detail.

\section{Bell, CHSH, GHZ}

Testing two-particle correlations in the Bell-CHSH setup \cite{bell,chsh}, as well as three-particle correlations in the GHZ setup \cite{ghz}, were a milestone in rendering quantum foundations a reliable object of research. The common scheme is to test entangled particles prepared in a reproducibly identical way, using a set of detectors in different settings. The assumption that identical preparation of particles implies identical values of hidden parameters determining measurement outcomes (``local realism'') leads to predictions which are in disagreement with experiments.

In our scheme that is the natural thing to happen.\footnote{This discussion has been inspired by correspondance with Lev Vaidman.} If hidden parameters are the initial states of detectors, then different detector settings imply different hidden parameters. To strengthen that statement, macroscopically different detector settings imply orthogonality of the corresponding detector quantum states - a trivial case of Anderson's ``orthogonality catastrophe'' \cite{PWA}; therefore the postulate of identical values of hidden parameters is not satisfied. Instead, statistical predictions of quantum mechanics are reproduced through the ``gambler's ruin'' dynamics described by Eq. (\ref{pump}), in agreement with experiments.

\section{No heating}

An interesting prediction of noise-based collapse models, heating due to explicit breaking of time translation symmetry by external white noise \cite{dlheat,noheat,adler}, is absent in the present scheme: noise generation being attached to strongly chaotic dynamics of detectors in the pre-firing period, nothing would break time translation symmetry, leaving no reason for heating in non-measurement situations. In particular, this should be true for the possibility of nucleon decay through bound state excitation \cite{pearlesquires,mayburov}.\footnote{I thank S. Adler for calling my attention to that work.} 

A formal argument to demonstrate that heating is absent in our approach is this: in the interaction picture, using cyclic invariance of the trace, the mean value of the Hamiltonian is
\be\label{noheating}
\langle \hat H(\hat{\boldsymbol\varrho}(t))\rangle~=~ i\zeta ~Tr 
      \left[ \left(\hat x(t) \hat{\boldsymbol\varrho}(t)\hat p(t) 
      - \hat p(t) \hat{\boldsymbol\varrho}(t) \hat x(t)\right)\hat{\boldsymbol\varrho}(t) \right] ~=~ 0
      ~~~~\forall t,
\ee
which excludes that energy, conserved in linear quantum dynamics, might be changed by nonlinearity.

\section{Locality and direct product structure}

Although the emblematic case of Bell-CHSH and GHZ correlations is handled by the above argument, more insight to what is local and what is not in the dynamics of a multi-detector setup can be obtained by analyzing the structure of solutions of Eq. (\ref{vNnonlinfinal}). 

The separability of the scalar product offers a convenient starting point to study the issue of locality, since the Hamiltonian governing the evolution according to Eq. (\ref{vNnonlinfinal}) is decomposed into a sum over detectors:
\be\label{sumoverdet}
\hat H(\hat{\boldsymbol\varrho}) = \sum_d  \left(\hat H_0^d  + i\zeta \left( \hat x_d \hat{\boldsymbol\varrho} \hat p_d -  \hat p_d \hat{\boldsymbol\varrho} \hat x_d \right)\right).
\ee
Here $\hat H_0^d$ describes coupled linear evolution of detector $d$, its interacting environment, as well as the respective partial wave of incoming particles initially interacting with that detector. As an immediate consequence, Eq. (\ref{nondiag}) takes the form
\be\label{nondiagsep}
\frac{\partial \hat R_{kl}}{\partial t}=~\frac{\zeta}{\hbar} ~\sum_d \sum_m w_m 
\left(\langle\hat p_d\rangle_m \cdot \{\hat x_d,\hat R_{kl}\} 
- \langle\hat x_d\rangle_m\cdot\{\hat p_d,\hat R_{kl}\}\right).
\ee
The pumping rate equation (\ref{pump}) remains unchanged, with Eq. (\ref{toandfro}) separating into the sum of one-detector terms:
\be\label{toandfrosep}
A_{km}(t)=~\frac{2}{\hbar}~\sum_d~  \Big(\langle\hat x_d\rangle_k \cdot \langle\hat p_d\rangle_m
             ~-~ \langle\hat x_d\rangle_m \cdot \langle\hat p_d\rangle_k\Big).
\ee

An important property of the relevant solutions of Eq. (\ref{nondiagsep}) emerges if we assume that before measurement starts, remote detectors are uncorrelated. That imposes that following interaction with the separated partial waves of incoming particles from a common source, the multi-detector density matrices $\hat R_{kl}$ in Eq. (\ref{dnstmtrx}) take the direct product structure 
\be\label{dirprod}
\hat R_{kl}(t) ~=~ \bigotimes_d \hat r_{kl}^d(t)
\ee
with $\hat r_{kl}^d=|\varphi_k^d\rangle\langle\varphi_l^d|$.\footnote{In Ref. \cite{jphysa18} the one-detector density matrix elements were assumed to be built of normalized one-detector states, varying local weights being incorporated into a single global one; that gave rise to ill-defined equations, as noted in the Corrigendum. In the present treatment the norms of each $|\varphi_k^d\rangle$ are changing by dynamics.}  In accordance with our expectations, that structure is preserved during subsequent evolution. Substituting Eqs. (\ref{dirprod}) and (\ref{sumoverdet}) into Eq. (\ref{vNnonlinfinal}), then following steps preceding Eq. (\ref{nondiag}), including use of the interaction picture, one obtains a sum over $d$ of the analogous one-detector equations
\be\label{onedet} 
\frac{\partial \hat r_{kl}^d}{\partial t} =\frac{\zeta}{\hbar}\sum_m w_m 
\left(\langle\hat p_d\rangle_m \cdot \{\hat x_d,\hat r_{kl}^d\} 
- \langle\hat x_d\rangle_m\cdot\{\hat p_d,\hat r_{kl}^d\}\right),
\ee
for each $d$ multiplied by a factor $\bigotimes_{d'\not=d}\hat r_{kl}^{d'}$. 

We introduce the one-detector weights $w_k^d(t)$ as
\be\label{detweight}
w_k^d(t)~=~Tr_d ~\hat r_{kk}^d(t),
\ee
connected to the global weights $w_k$ through Eqs. (\ref{dirprod}) and (\ref{weight}) as
\be\label{proddyn}
w_k(t)~=~|c_k|^2~\prod_d w_k^d(t);~~~
    \frac{\dot w_k}{w_k}~=~\sum_d~\frac{\dot w_k^d}{w_k^d}.
\ee
Using Eqs. (\ref{onedet}) and (\ref{detweight}), each term on the r.h.s. obeys a one-detector dynamical equation
\be\label{detweightdyn}
\frac{\dot w_k^d}{w_k^d}~=~\frac{2\zeta}{\hbar}~\sum_m~ w_m \Big(\langle\hat x_d\rangle_k \cdot \langle\hat p_d\rangle_m ~-~ \langle\hat x_d\rangle_m \cdot \langle\hat p_d\rangle_k\Big).
\ee
Finally, we see that the pumping equations (\ref{pump}) to (\ref{antisymm}) remain valid, however, the product structure expressed in Eq. (\ref{proddyn}) implies a strange combination of single-detector dynamics: $w_k\to 0$, as soon as $w_k^d\to 0$ at {\sl anyone} of the detectors. This assigns an unexpectedly strong meaning to the short-living one-detector weights $w_k^d$, to be explored by further research. Further insight is obtained by noticing that for any operator $\hat A_d$ acting on the configuration subspace of detector $d$, like $\hat x_d$ and $\hat p_d$, in Eq. (\ref{meanink}) the $d'\not=d$ multipliers cancel, therefore all one-detector averages appearing in Eq. (\ref{detweightdyn}) are fully determined locally, and can be calculated by $\hat r_{kk}^d$:
\be\label{meaninkd}
 \langle \hat A_d \rangle_k
 ~=~\frac{\langle\varphi_k|\hat A_d |\varphi_k\rangle}{\langle\varphi_k^d|\varphi_k^d\rangle}
   ~=~\frac{Tr_d\left(\hat r_{kk}^d\hat A_d\right)}{Tr_d~\hat r_{kk}^d}.
\ee

As already emphasized in Ref. \cite{jphysa18}, it is important to note that whereas Eqs. (\ref{onedet}) are quasi-local, since each of them refers to a single detector $d$, they include the global weights $w_m$, representing remote entanglement in the scenario. As mentioned in the Introduction, this does not allow signaling through statistics, as granted by the results summarized in Ref. \cite{brunner}: since our treatment reproduces Born's-rule-based quantum statistical predictions, it remains a subset of non-signaling theories. Thereby, the performance of our model is equivalent to that of CSL-type models \cite{csl}, with the advantage that here noise appears as a consequence of multilocal dynamics of the measurement apparatus. However, "spooky action-at-a-distance"  in individual events, not accessible through measurement statistics but still part of reality, remains with us. 

\section{Final remarks}

As discussed in more detail in Ref. \cite{jphysa18}, our treatment demonstrates that a deterministic, nonlinear dynamical equation can be capable to reproduce Born's rule quantum statistics, without signaling. Unlike CSL-type theories \cite{csl}, no external noise is needed for that, randomness emerging from chaotic dynamics. That is demonstrated for genuine measurement situations; in non-measurement contexts different statistics, eventually, 1/f noise \cite{oneoveref} can emerge.

Nonlinearity challanges our feeling of comfort offered by linear quantum mechanics. In particular, the equivalence of averaging a density matrix over a Gibbs ensemble of different states, and obtaining the same density matrix by tracing out over a big environment, is lost by nonlinearity. In the present case, it is easy to check that the derivation of Born's rule is insensitive to that difference; however, fine details of collapse dynamics, as well as some details of non-measurement dynamics, can show up observable consequences of the loss of that linearity-based equivalence.

{\bf Acknowledgment.} It is my pleasure to acknowledge endless discussions over many years with Lajos Di\'osi; special thanks for a critical reading of an earlier version of the present manuscript. I thank helpful recent correspondance with Philip Pearle, Lev Vaidman and Stephen Adler, and stimulating discussions with Michael Vanner at DICE2018. 


\end{document}